\documentclass[a4paper,12pt,twoside]{article}
\usepackage{graphicx}
\usepackage{amsmath}
\usepackage{amssymb}
\usepackage{eufrak}
\usepackage{color}
\usepackage{enumerate}
\usepackage{dsfont}
\def\singlesided{\evensidemargin=0pt}
\singlesided

\numberwithin{equation}{section}
\begin{document}
\begin{verbatim}

\end{verbatim}

\centerline{\Large \bf Towards Quantum Field Theory in Curved Spacetime}
\vspace{0.1cm}
\centerline{\Large \bf for an Arbitrary Observer}

\vskip 1.5\baselineskip
\centerline{\large Hui Yao}

\vspace{0.2cm}
\centerline{\texttt{hy255@cam.ac.uk}}

\vspace{0.2cm}
\centerline{ \textit{DAMTP, University of Cambridge}}

\begin{verbatim}
\end{verbatim}

\thispagestyle{empty}
\begin{abstract}

We propose a new framework of quantum field theory for an arbitrary observer in curved spacetime, defined in the spacetime region in which each point can both receive a signal from and send a signal to the observer. Multiple motivations for this proposal are discussed. We argue that radar time should be applied to slice the observer's spacetime region into his simultaneity surfaces. In the case where each such surface is a Cauchy surface, we construct a unitary dynamics which evolves a given quantum state at a time for the observer to a quantum state at a later time. We speculate on possible loss of information in the more general cases and point out future directions of our work.
 
\end{abstract}
\catcode`\@=11
\c@page=3
\catcode`\@=12
\begin{verbatim}
\end{verbatim}
\tableofcontents
\vfil\eject
\pagestyle{plain}
\catcode`\@=11
\c@page=0
\catcode`\@=12
\pagenumbering{arabic}

\def\bb{\begin{equation}}
\def\ee{\end{equation}}
\def\cc{\cite}
\def\etal{\textit{et al. }}

\def\hh{\mathcal{H}}
\def\ff{\mathcal{F}_s}
\def\ccc{\mathcal{C}(t)}
\def\ccci{\mathcal{C}(t_1)}
\def\cccii{\mathcal{C}(t_2)}
\def\aa{\mathcal{A}}

\section{Motivation}

The Universe is ultimately observer-participatory, each part of which is communicating with the others and never at rest. Existence are not simply ``things just out there'', but reality for an observer is his dynamical construction. Both quantum theory and relativity, the two pillars of our modern conceptual understanding of Nature, taught us so, although the great union of the two remains a mystery still. We shall hence aim at a further understanding of the observer's quantum description of matter in a general relativistic background. We first discuss the multiple motivations of this paper.

\begin{enumerate}[\textbullet]
\item Quantum field theory in curved spacetime is the necessary first step towards a conceptual understanding of a quantum theory of gravity. Even though such a framework is ultimately incomplete, we expect some features of the theory to remain. We shall treat the quantum field like a ``test field'' with the background spacetime completely classical and unaffected by the matter field.

\item We view physics as a theory of an arbitrary observer's dynamical description of his system. Therefore any physical theory should be formulated explicitly in terms of an observer's own physical quantities at a fundamental level. There have been great conceptual advancements in understanding the observer-dependent aspects of quantum field theory [1-6]. Rather than treating observer-dependence as an emergent feature, we now push these developments further by taking observer-dependence to be a cornerstone of our formulation of quantum field theory in curved spacetime.

\item Working at a semi-classical level, we shall understand an observer as a history of spacetime events, in other words, a timelike curve in a classical curved spacetime. A quantum state is a quantum description of the physical system for the observer at some proper time of his. We consider dynamics as the updating of his description of the system rather than as the ``changing of things in themselves''. Above all, we would like to construct a theory specifying a one-parameter family of quantum states along a timelike worldline. 

\item Dynamics is the specification of a one-parameter family of physical states, whether they be points in classical phase spaces or vectors in Hilbert spaces or some other mathematical representation. Therefore intrinsic to dynamics is a notion of ``time''. A notion of time is usually only available in special cases. One may restrain oneself to the notion of asymptotic past/future only, but this does not define a parameter of physical time. There is a ``natural'' choice of time if the spacetime manifold admits a special symmetry, like the Killing time for stationary spacetimes or the usual cosmic time of Friedmann-Robertson-Walker (FRW) cosmology. However, the last two choices are rather mathematical and only apply to special cases. We would instead like a single physically-motivated definition applicable to all situations. The only possibility of such a definition is the proper time on an observer's clock. Such a choice in fact forces one to adopt an observer-dependent description of physics, which is precisely what we have been aiming for.
   
\item In this paper we shall take a Hilbert space approach in that we consider Hilbert spaces to be the most fundamental objects mathematically representing our quantum description of physical systems. Quantum states are vectors or more generally density matrices in these spaces. Such a mathematical framework is especially suitable for discussing quantum state evolution, where a two-parameter mapping $U(t_2, t_1)$ relates a quantum state in a Hilbert space at some time $t_1$ for an observer to another state in a possibly different Hilbert space at time $t_2$. The Hilbert space approach is also particularly suitable for discussing any possible loss of information in the form of a pure state evolving to a mixed density matrix.

\item There are two important conceptual lessons that we should draw from Hawking's semi-classical analysis \cc{Hawking:1976, Hawking:1982} of black hole information loss. Firstly, Hawking has insisted that any quantum state for the observer outside the black hole must be physically meaningful for him: to obtain the correct description of the quantum system according to the observer, we must trace out all degrees of freedom that he cannot causally access. As we shall see, our quantum theory is precisely defined over the spacetime region causally connected to the given observer. Secondly, that there is a loss of information when black hole evaporates indicates that the loss of information is rooted in the ``evolution'' of a horizon rather than simply the presence of a horizon. To make precise sense of evolution one again needs a notion of time, which we have chosen to be the proper time of an observer. 

\item We shall speculate that there will be a non-unitary evolution and hence information loss for an observer, precisely when his surface of simultaneity evolves from a Cauchy surface to a non-Cauchy surface for the spacetime region to which the observer has causal access. Instead of resolving information loss as a paradox, we propose to forcefully carry Hawking's argument \cc{Hawking:1976, Hawking:1982} through and speculate that the possibility of information loss is a fundamental feature of quantum field theory in curved spacetime rather than special to black hole evaporation.
 
\item  To obtain physical quantities from the point of view of an observer, one normally has to study the response of a model particle detector following the observer's worldline. We would instead like to have a fundamental theory in which the particle content at any time can be simply read off from the mathematical representation of the physical state. Furthermore a quantum state, which completely encodes all information about the possible measurement outcomes and their probabilities of occurrence, tells much more than simply the particle spectrum, in particular whether the quantum state is pure or mixed.

\item Wald \cc{wald} has forcefully argued that there is no natural choice of Fock space of particles in quantum field theory in a most general spacetime, and that this is in analogy to choosing a coordinate system on a manifold in general relativity which cannot be physical. But for a given observer, as we shall show, there indeed exists a natural notion of ``particles''. There have been serious difficulties in attempting to define a notion of particles in a most general spacetime, but this does not imply that the idea of ``particle'' itself is not fundamental, as long as one takes an observer-dependent viewpoint of quantum field theory. 

\item A notion of particles is usually only available in special cases such as if the spacetime admits a special symmetry or special asymptotic behaviours, or if the physical situation admits an adiabatic approximation. However generalisation is essential, as elements particular to special cases can obscure what the fundamental features are. The formulation which we will present can be applied rather generally to a wide class of observers without requiring any special symmetries or asymptotic behaviours of the spacetime. 

\item Unlike in e.g. \cc{Hajicek}, here we shall make no fundamental distinction between particle creation due to the motion of the observer and that due purely to gravitational fields. We view spacetime as a geometric and causal background in which the observer's worldline is defined. All observers are regarded as completely equivalent at the fundamental level of quantum field theory in curved spacetime.

\item Ashtekar and Magnon \cc{am} have constructed a one-parameter family of Fock spaces in their formulation of quantum field theory in curved spacetime. The authors found their construction to depend on a choice of timelike vector field and hence of the corresponding integral curves. They were therefore forced to conclude that their theory depends on a field of observers. Rather than be led to a conclusion of observer-dependence, we have taken as our conceptual starting point the construction of  a quantum field theory with observer-dependence.

\item One might interpret Ashtekar and Magnon's construction \cc{am}, which depends on a congruence of timelike curves, to be for a family of observers. However, we insist any physical theory should be formulated for an arbitrary single observer. Mathematically, as we shall show, the construction of the family of Fock spaces depends only on a choice of scalar function $t$ and a corresponding foliation. Physically, if one starts with a single observer, in a most general situation it is far from clear how to choose a family which he is a member of.  Furthermore, specifying an arbitrary family of observers has no direct physical interpretation as one can never set up an experiment with an uncountably many number of detectors which trace out a congruence of curves covering the entire spacetime. Finally, one expects that different observers within one family, even if there is such a preferred grouping, would differ in their description of a physical system. One therefore would like this difference to be naturally accounted for and built into the fundamental theory.

\item This work in some sense parallels Einstein's construction of special relativity. Just as the principle of relativity is the guiding principle of special relativity, it is our conceptual starting point that the same laws of quantum field theory should apply to any arbitrary observer, although the observers' dynamical quantum descriptions may differ. Secondly, just as Einstein recognised the inseparable connection between time and the signal velocity, we shall apply radar time, which is operationally defined using light signal communication, to formulating quantum field theory for an arbitrary observer in a general spacetime.
\end{enumerate}

The plan of the paper is given as follows.

In section 2, we shall construct a one-parameter family of Fock spaces based on the formalism of Ashtekar and Magnon \cc{am}. We shall show that the formalism requires a choice of scalar function $t$. 

In section 3, we shall apply radar time to operationally define this function $t$ for each point which a given observer can both send a signal to and receive a signal from. We regard defining quantum field theory in the spacetime region causally connected to a given observer as an axiom of our framework. In section 3.1 we discuss the importance of Cauchy surfaces in the formulation of a unitary theory and the preservation of information.

In section 4, we construct the dynamics of quantum states in our theory. We define quantum evolution in terms of a two-parameter mapping from one Fock space to another, each associated with a time for an observer, and we show this mapping satisfies certain necessary physical conditions. 

Finally in section 5, we summarise our main results and consider possible directions in which our work might be developed.

We shall use natural units throughout this paper. The sign convention in general relativity is the same as that of \cc{MTW}, in particular, $\eta_{ab}=(-1,+1,+1,+1)$.

\section{The Hilbert Spaces of the Free Real Scalar Field}

In this section, we shall construct a one-parameter family of Fock spaces of the free real scalar field based on the work of Ashtekar and Magnon \cc{am}. Definitions and notations are introduced which will be used throughout the subsequent sections. We shall summarise the formalism in a form most simple and ready for our purpose of the Hilbert space approach as motivated in section 1. In particular, we shall not start from the *-algebra of abstract field operators of \cc{am}. 

Let $V$ be the vector space of all well-behaved\footnote{Ashtekar and Magnon \cc{am} have assumed that all solutions in $V$ are smooth and induce, on any spacelike Cauchy surface, initial data sets of compact support. The assumption about compact support is used to ensure convergence of various integrals and to discard various surface terms when integrating by parts.} real-valued solutions of Klein-Gordon equation
\bb
(\nabla^a\nabla_a - m^2)\,\phi = 0
\ee
on a given globally hyperbolic spacetime. Let $\Sigma$ be an arbitrary spacelike Cauchy surface of the given spacetime, with arbitrary coordinates $\{x^i\}$ and unit future-directed normal $n^a$. The induced metric on $\Sigma$ is $h_{ab}$. A symplectic form $\omega$ on $V$ is defined as
\bb
\omega\,(\phi,\psi) = \displaystyle{\int_{\Sigma}}(\psi\nabla_a\phi - \phi\nabla_a\psi)\,n^a\sqrt{h}\,d^3x.
\label{symplectic}
\ee
Let $J$ be a complex structure on the real vector space $V$, i.e. an automorphism on $V$ which satisfies $J^2 = -\mathds{1}$. $J$ endows $V$ with the structure of a complex vector space, which we shall denote as $V_J$. We shall use $|\;\rangle$ to distinguish an element $|\phi\rangle$ in $V_J$ from its counterpart $\phi$ in $V$. Hence in our notation we have, for example, $i|\phi\rangle = |J\phi\rangle$. An inner-product $\langle \; | \; \rangle$ can be defined on $V_J$ as
\bb
\langle \phi | \psi \rangle = \tfrac{1}{2}\,\omega(\phi,J\psi) + \tfrac{i}{2}\,\omega(\phi,\psi).
\label{inner_prod}
\ee
This indeed defines an inner-product if and only if the complex structure $J$ is compatible with the symplectic form $\omega$, i.e.
\bb
\omega\,(\phi,\psi) = \omega\,(J\phi,J\psi), \qquad \forall \phi,\,\psi \in V \label{symplectic1}
\ee
\bb
\omega\,(\phi,J\phi) > 0 \qquad \forall \phi \in V \backslash \{0\}.\label{symplectic2}
\ee
The Cauchy completion of the complex inner-product space $(\,V_J, \langle\;|\;\rangle\,)$ is a Hilbert space which we shall denote as $\hh_J$.

We now define the $n$-particle space to be the Hilbert space $\otimes^n_s\,\hh_J$, i.e. the $n$th-rank symmetric tensor over $\hh_J$. The space of all quantum states is then the symmetric Fock space $\ff(\hh_J)$ based on the Hilbert space $\hh_J$:
\bb
\ff(\hh_J) = \oplus_{n=0}^{\infty}\!\otimes^n_s\,\hh_J.
\ee
Creation and annihilation operators are defined as mappings on this $\ff(\hh_J)$ in the usual way; see e.g. \cc{wald}. We shall denote the creation and annihilation operators associated with $|\phi\rangle$ as $C^\mu(\phi)$, where $\mu= +1$ is for creation and $\mu = -1$ is for annihilation.\footnote{For typographical convenience, we did not choose the notation of $C^\mu(|\phi\rangle)$. Although it should be understood that $C^\mu(\phi)$ depends on $|\phi\rangle \in \hh_J$ rather than on $\phi \in V$.} The use of the index notation $\mu$ will become clear in section 4. One can show from their definitions these operators satisfy the following properties: (i) $(C^\mu(\phi)\,)^\dag = C^{-\mu}(\phi)$; (ii) each creation/annihilation operator is complex-linear/anti-linear in its argument; and (iii)
\bb
[C^+(\phi), \, C^+(\psi)] = [C^-(\phi), \, C^-(\psi)]  = 0, \quad [C^-(\phi), \, C^+(\psi)] = \langle \phi|\psi \rangle\, \mathds{1}.
\label{commutations}
\ee

To summarise, we have constructed a Fock space for each choice of a complex structure $J$ on $V$ which is compatible with the symplectic form $\omega$ in the sense of (\ref{symplectic1}) and (\ref{symplectic2}). To finish the construction, it remains to specify a one-parameter family of complex structures $J_t$ satisfying (\ref{symplectic1}) and (\ref{symplectic2}).

We now summarise, in a slightly different form,  the construction of $J_t$ due to Ashtekar and Magnon \cc{am}. Let $t$ be a scalar function on the spacetime such that each constant $t$ hypersurface $\Sigma_t$ is a spacelike Cauchy surface and the set $\{\Sigma_t\}$ foliates the given spacetime. Let $n^a = N^{-1} \left(\tfrac{\partial}{\partial t} \right)^a$ be the unit future-directed normal to $\Sigma_t$, where 
\bb
N = \sqrt{-g_{ab} \, \left(\tfrac{\partial}{\partial t}\right)^a \left(\tfrac{\partial}{\partial t}\right)^b}.
\ee 
To construct $J_t$, we introduce a $t$-dependent Hamiltonian operator $H_t$ on $\hh_t$ defined by\footnote{The subscript $t$ on $\hh_t$ is to remind us that it is constructed out of $J_t$ as described previously.}
\bb
H_t\,|\phi\rangle_t := -i\,|\tilde{H}_t \,\phi\rangle_t = -|J_t\, \tilde{H}_t \, \phi\rangle_t
\ee
where $\tilde{H}_t$ is a real-linear operator on $V$ which is defined as follows. If $\phi$ has on $\Sigma_t$ the Cauchy data $\phi|_t = f$, $n^a \nabla_a \phi|_t = g$, then $\tilde{H}_t \, \phi \in V$ is the solution with Cauchy data $\tilde{H}_t \, \phi|_t = Ng$, $n^a \nabla_a (\tilde{H}_t \, \phi)|_t = -N^{-1} \Theta f$ on the same Cauchy surface, where
\bb
\Theta := -N^2 h^{ab} D_a D_b - N h^{ab} (D_a N) D_b + m^2 N^2;
\ee 
$h_{ab}$ is the induced metric on $\Sigma_t$; and $D_a$ is the covariant derivative on $(\Sigma_t,\, h_{ab})$. It is easy to see that $H_t$ is complex-linear on $\hh_t$ if and only if $\tilde{H}_t$ is a real-linear operator on $V$ and $\tilde{H}_t$ commutes with $J_t$, i.e. $[\tilde{H}_t,\, J_t] = 0$. 

It is proven \cc{am} that there exists a unique complex structure $J_t$ satisfying (\ref{symplectic1}) and (\ref{symplectic2}) such that $\langle \phi | H_t | \phi \rangle_t$ is real for any $|\phi\rangle_t \! \in \! \hh_t$. If $\phi$ is the solution with Cauchy data $(f, \, g)$ on $\Sigma_t$ as defined before, then $J_t \, \phi$ is the solution with Cauchy data $(\Theta^{-\frac{1}{2}} N g, \, -N^{-1} \Theta^{\frac{1}{2}}f)$ on the same Cauchy surface. One can check that $J_t$ does indeed commute with $\tilde{H}_t$: $[J_t, \tilde{H}_t] = 0$. Furthermore, the following relation holds automatically
\bb
\langle \phi | H_t | \phi \rangle_t = \displaystyle{\int_{\Sigma_t}} T_{ab}\,N\,n^a\,d\Sigma^b,
\ee
where the energy-momentum tensor $T_{ab}$ is given by
\bb
T_{ab} = \nabla_a \phi \nabla_b \phi - \tfrac{1}{2} g_{ab} (\nabla^c \phi \nabla_c \phi + m^2 \phi^2).
\ee

In applying the above formalism to obtain the quantum theory as motivated and outlined in section 1, we need to solve two more problems.

First of all, we would like to interpret the above mathematical formalism physically for an arbitrary single observer in a given spacetime. In the above construction of the one-parameter family of Fock spaces, there is an ambiguity in the choice of the scalar function $t$. We need to specify the scalar $t$ and understand its operational meaning for a given observer. This we shall discuss in the next section.

Secondly, we would like to understand the dynamics of the theory. A differential form of the dynamics has been discussed in \cc{am}, however we will not follow that approach. The question we would like to answer is: given a quantum state for an observer at his proper time $t_1$, what is the evolved quantum state at another time $t_2$? In other words, we need to construct a two-parameter mapping from $\ff(\hh_{t_1})$ to $\ff(\hh_{t_2})$ satisfying certain properties, which we shall discuss in section 4.

\section{Simultaneity for an Arbitrary Observer}
In the previous section, we have constructed a one-parameter family of Fock spaces. The construction relies on a choice of scalar function $t$ and a corresponding slicing of the spacetime $\{\Sigma_t\}$. In this section, we would like to understand this $t$ for a spacetime event $p$ as the ``time'' of $p$ for a given arbitrary observer, and $\Sigma_t$ as the set of all events ``simultaneous'' to the observer at his proper time $t$. We would like to understand how the events of a given spacetime are directly related to the local physical quantities of the observer in an operational way. 

To understand the physical meaning of this ``time'' and ``simultaneity'', we first look at an inertial observer $\ccc$ in Minkowski spacetime. Let $t_p$ be the earliest time when $\ccc$ can receive a signal from an event $p$. Let $t_p^\prime$ be the latest time for $\ccc$ to send a signal to reach $p$. We define $t_q$ and $t_q^\prime$ analogously for another event $q$. Then the events $p$ and $q$ are simultaneous for the inertial observer $\ccc$ if and only if the relation $t_p-t_q=t_q'-t_p'$ holds. This captures the intuition that if $p$ and $q$ are simultaneous then the difference in inquiring time and response time would equal, and this difference is purely due to any difference in the distances from $p$ and $q$ to the observer. This construction, known as \textit{radar time}, has been advocated by Bondi \cc{bondi}. Applications of radar time to an observer-dependent particle interpretation in quantum field theory has recently\footnote{I would like to thank David Wiltshire for letting me be aware of \cc{gull} after completion of a draft of this paper.} been pioneered by Dolby and Gull \cc{gull}.

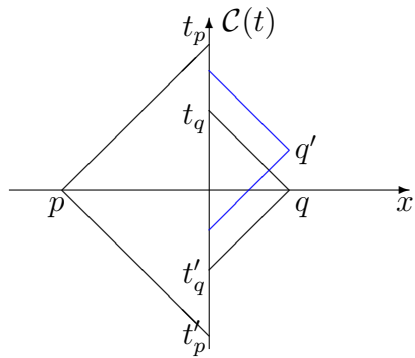
\begin{figure}[!h]
\begin{center}
  \begin{picture}(150,130)
    \put(0,65){\vector(1,0){150}}
    \put(75,5){\vector(0,1){125}}
    \put(20,65){\line(1,1){55}}
    \put(20,65){\line(1,-1){55}}
    \put(105,65){\line(-1,1){30}}
    \put(105,65){\line(-1,-1){30}}
    \textcolor{blue}{\put(105,80){\line(-1,1){30}}
	\put(105,80){\line(-1,-1){30}}}
    \put(15,57){$p$}
    \put(107,57){$q$}
    \put(107,77){$q^\prime$}
    \put(145,57){$x$}
    \put(80,125){$\ccc$}
    \put(65,123){$t_p$}
    \put(65,7){$t_p^\prime$}
    \put(65,90){$t_q$}
    \put(65,30){$t_q^\prime$}
  \end{picture}
\caption{The events $p$ and $q$ are simultaneous, but $p$ and $q^\prime$ are not.}
\end{center}
\end{figure}

The above characterisation of simultaneity is readily generalisable to an arbitrary observer in a curved spacetime. Let an observer $\ccc$ be a timelike curve parametrised by his proper time $t$ and consider a spacetime event $p$. Let $\tau_p$ be the earliest\footnote{It may happen that a future-directed null geodesic originating from a point $p$ intersects an observer's worldline multiple times. For example, in cylindrical spacetime with metric $ds^2 = -dt^2 + d\theta^2$, a null geodesic from a point $p$ intersects a constant $\theta$ observer infinitely many times. In this case, our definition of $\tau_p$ is the first future point of intersection. Similar remarks hold for $\tau_p^\prime$.} proper time of the observer such that there is a future-directed null geodesic originating from $p$ to $\mathcal{C}(\tau_p)$, and let $\tau_p^\prime$ be the latest time such that there exists a future-directed null geodesic originating from $\mathcal{C}(\tau_p^\prime)$ to $p$. If there exist both $\tau_p$ and $\tau_p^\prime$, then we say that $p$ is in causal contact with the observer.

Following \cc{diamond}, we shall call the set of all points in causal contact with the observer $\ccc$ the \emph{diamond} of $\ccc$. The diamond is thus defined to be the set of all points which can both receive light signals from and send light signals to the observer $\ccc$, in other words, those points which can exchange information with $\ccc$.  Equivalently, the diamond is defined to be the intersection of the past of $\ccc$ and the future of $\ccc$. We may also talk about the horizon for an observer $\ccc$, most generally, as the boundary of the diamond of the observer.

We define the scalar function $t(p)$ for each point $p$ in the diamond of an observer to be
\bb
t(p) := (\tau_p + \tau_p^{\prime})\,/2,
\label{prop_time}
\ee
and we say $p$ is simultaneous to the observer at $t(p)$. In the case when the point $p$ is lying on $\ccc$, i.e. $p = \mathcal{C}(t_0)$ for some $t_0$, we define $t(p)=t_0$. Given a time $t_0$, we define the \emph{surface of simultaneity} $\Sigma_{t_0}$ to be the set of all points $p$ in the diamond for which $t(p) = t_0$. Since each point $p$ of the diamond belongs to exactly one $\Sigma_{t(p)}$, we obtain a foliation $\{\Sigma_t\}$ of the entire diamond. Furthermore, we have a timelike vector field $\partial / \partial t$ everywhere orthogonal to $\{\Sigma_t\}$.

The simultaneity of our definition depends on the observer's entire history; that is, the surface of simultaneity $\Sigma_{t_0}$ at time $t_0$ depends on the observer's trajectory $\ccc$ for all $t$. This should not be taken as a flaw of our definition. Firstly, our definition using light signal communication has a clear operational meaning. Secondly, in Minkowksi spacetime with the usual coodinates $(t, x^i)$, the constant $t$ surfaces are the usual planes of simultaneity for an inertial observer with constant $x^i$, and the specification of the entire trajectory of constant $x^i$ is essential. Thirdly, if two observers locally coincide around a common point, their surfaces of simultaneity through that point will also be locally the same.

One could instead naively define simultaneity surface $\Sigma_{t_0}$ to be the set of all points lying in geodesics orthogonal to the observer's trajectory $\ccc$ at $t_0$. This definition of $\Sigma_{t_0}$ depends only on the point $\mathcal{C}(t_0)$ and its tangent vector, and for an arbitrary observer $\ccc$ in Minkowski spacetime it gives the usual simultaneity plane for the inertial observer tangent to $\ccc$ at $\mathcal{C}(t_0)$. However, the physical meaning of these surfaces for an arbitrary observer becomes unclear far from the worldline. Furthermore, these surfaces may intersect \cc{MTW} within the observer's diamond, so that  the ``time'' of the point of intersection becomes ambiguous under this definition. Our definition of simultaneity does not suffer from the same problem, as each point in the diamond of the observer belongs to exactly one simultaneity surface.

We have defined a scalar function $t$ and a corresponding foliation $\{\Sigma_t\}$ of the diamond region of an arbitrary observer. Application of this idea to the formalism summarised in section 2 implies that we should treat the diamond as the physically relevant part of spacetime for the observer, on which the vector space $V$ of real Klein-Gordon solutions is defined. 

Operationally, only those events in the diamond of an observer are physically relevant to him. If there exists a future-directed null geodesic originating from a point in $\ccc$ to a point $p$, then one may naively say that the observer can send a signal to $p$. However this has no causal effect on him if there is no future-directed null geodesic from $p$ to $\ccc$. The only way that the observer is able to operationally determine whether $p$ has received his signal is if in principle he can receive a reply from $p$. Similarly, any event $q$ to which there is no future-directed null geodesic from $\ccc$ connecting is physically irrelevant.

It is in fact a more or less usual practice in quantum field theory that one interprets only the degrees of freedom in the region causally connected to the observer as the physically relevant ones. In the example of the Rindler observers, only the field in the Rindler wedge is considered relevant \cc{Unruh}. For observers staying outside of a black hole, only the field outside the black hole horizon is relevant \cc{Hawking:1976, Hawking:1982}. And for an inertial Minkowski observer of a finite-lengthed worldline, only the field in the observer's diamond region is relevant \cc{diamond}. Rather than interpreting the physically relevant spacetime region for a given observer on a case by case basis, we here regard defining quantum field theory on only the diamond for an arbitrary observer as an axiom of our framework.
\subsection{Cauchy surfaces and information loss}

The formalism in section 2 requires that each surface $\Sigma_t$ of the foliation be a Cauchy surface. Our slicing of the diamond for an arbitrary observer defined earlier in this section, however, does not sastisfy this requirement in general. It may indeed be the case that all simultaneity surfaces of the observer are Cauchy surfaces for his diamond, as in the case of the Rindler observer. But it may also be that at least one but not all simultaneity surfaces are Cauchy surfaces, as in the case of an inertial Minkowski observer with a finite worldline \cc{diamond}; or that none of the simultaneity surfaces are Cauchy surfaces, such as a co-moving observer in the Milne universe, even though the spacetime is globally hyperbolic. 

In the next section, we shall construct a unitary quantum theory in the case that all simultaneity surfaces are Cauchy surfaces. The construction for the more general cases has to remain future work. However, we argue here that the limitation of the current formulation should not be regarded as a fundamental flaw of our conceptual approach. We instead interpret this restriction as revealing the importance of Cauchy surfaces in the formulation of a unitary theory and the preservation of information. We shall now speculate on the qualitative features of the theory for the more general cases. 
 
Let $t_1$ and $t_2 $ be an observer's proper time with $t_2 > t_1$. Generally, when surface $\Sigma_{t_2}$ is not contained in the future Cauchy development $D^+(\Sigma_{t_1})$ of surface $\Sigma_{t_1}$, one cannot determine the data on $\Sigma_{t_2}$ and hence the physical state at $t_2$ from the data on $\Sigma_{t_1}$, and therefore the dynamics of the theory is non-deterministic. Conversely, if all of $\Sigma_{t_2}$ lies in $D^+(\Sigma_{t_1})$ but not all of $\Sigma_{t_1}$ lies in $D^-(\Sigma_{t_2})$, where $D^-(\Sigma_{t_2})$ is the past Cauchy development of $\Sigma_{t_2}$, then evolution of the field from $\Sigma_{t_1}$ to $\Sigma_{t_2}$ is deterministic but non-invertible and hence there is a loss of information. There will exist timelike curves crossing $\Sigma_{t_1}$ but not $\Sigma_{t_2}$, which physically means that there will exist worldlines of objects leaving the observer's horizon before he reaches $t_2$.

In the case where only one of the simultaneity surfaces $\Sigma_{t_0}$ is a Cauchy surface, such as the example in \cc{diamond}, one can determine the field throughout the entire diamond from $\Sigma_{t_0}$. One can predict the future of $\Sigma_{t_0}$, but information is lost as the field evolves from $\Sigma_{t_0}$ to a future simultaneity surface. One can also ``postdict'' at $t_0$ what happened before that time, although before $t_0$ he cannot predict what happens at $t_0$: there is more for him to learn, but in retrospect nothing is surprising.

We now consider the implications of the above discussion for the example of a gravitationally collapsing black hole without evaporation. As shown in Figure 2 (a), an arbitrary observer that does not fall into the black hole has a diamond corresponding to spacetime outside the black hole horizon. If the causal structure and the simultaneity surfaces are correctly depicted in our figure where all simultaneity surfaces are Cauchy surfaces, then we see that the observer's quantum field theory is unitary and that no information is lost. This is consistent with the intuition that an observer remaining outside the black hole never in his finite proper time sees an object falling behind the horizon and no information carried by the in-falling object could ever disappear from the sight of the observer. Even though the observer cannot access spacetime behind the horizon, that region is physically irrelevant to him: he does not know everything, but at least he knows what he knew.

\begin{figure}[htb]
\vbox{\vskip 10 pt
\centerline{
\begin{tabular}{c c c c c}
\includegraphics[width=50mm]{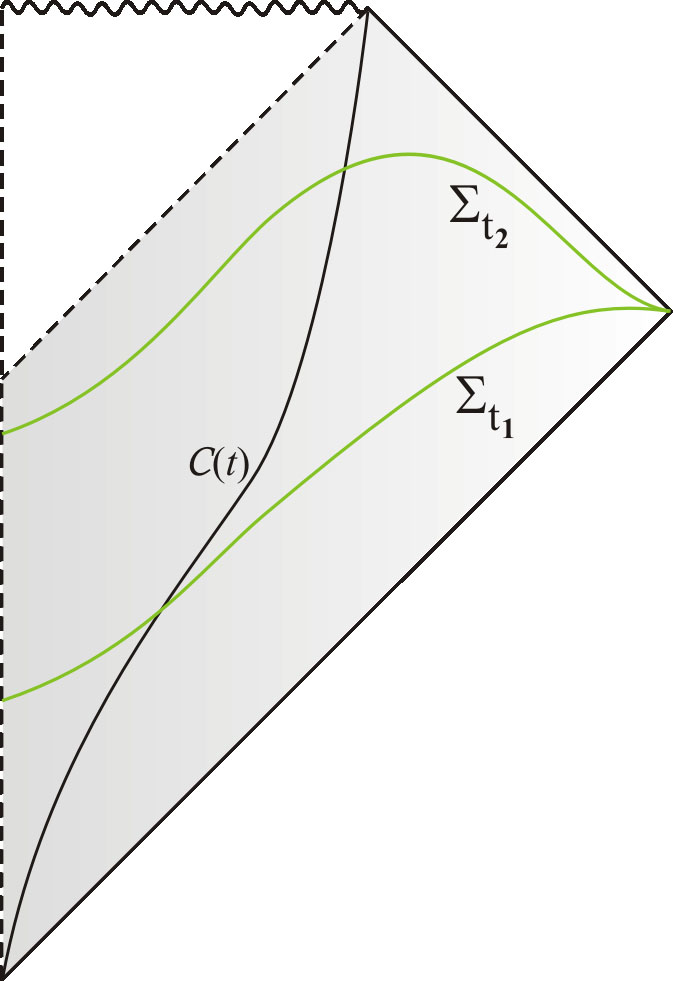}
&&&&
\includegraphics[width=50mm]{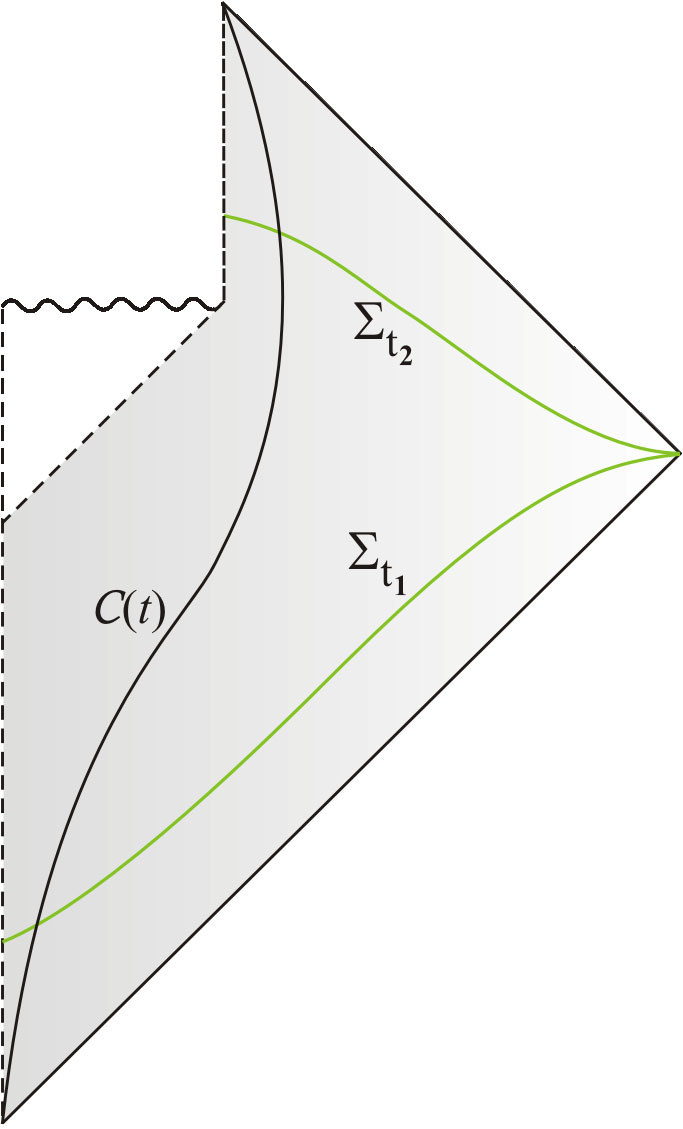}\\
(a) &&&& (b)
\end{tabular}}
\caption{\sl Diagrams of black hole spacetimes: (a) without evaporation; and (b) with evaporation. Two typical simultaneity surfaces $\Sigma_{t_1}$, $\Sigma_{t_2}$ for the observer $\ccc$ are shown. \label{bh}}}
\end{figure}

In the spacetime of black hole with evaporation, however, we speculate that not all simultaneity surfaces will be Cauchy surfaces for the diamond of the observer staying outside of the black hole, as depicted in Figure 2 (b). We expect that in the figure, $\Sigma_{t_2}$ will lie in $D^+(\Sigma_{t_1})$, but $\Sigma_{t_1}$ will not be contained in $D^-(\Sigma_{t_2})$.  Hence it follows that evolution will be deterministic, but information will be lost, a profound and long renowned proposal due to Hawking \cc{Hawking:1976, Hawking:1982}.

Wald \cc{wald} has already pointed out that there should generally be a loss of information associated with evolution of a Cauchy surface to a non-Cauchy surface. Here we have further developed the idea. With our one-parameter family of hypersurfaces $\{\Sigma_t\}$ constructed previously, we are able to provide Wald's ``evolution of surfaces'' a precise mathematical meaning. Physically, we give these ``surfaces'' an operational meaning as the simultaneity surfaces of a given observer. Moreover, the question of whether or not a surface is a Cauchy surface is now addressed with respect to the observer's diamond. We have emphasised that the observer should play a fundamental role in the discussion of a possible non-unitary quantum theory. 

We have proposed that Hawking's original insight that the laws of physics may allow for loss of information can be much generalised. Information can be lost not only in black hole evaporation but precisely whenever one's surface of simultaneity evolves from a Cauchy surface to a non-Cauchy surface, whether this be due to the background spacetime or to the observer's motion in that background. We emphasise that because our formulation reduces to the standard quantum field theory for inertial observers in Minkowski spacetime, allowing for a non-unitary theory for general observers in curved spacetime cannot be in violation of the usual laws of physics in flat spacetime. And as we have argued that quantum field theory should be formulated in an observer-dependent way, the notion that information loss is observer-dependent does not mean that it is not fundamental.

Here we do not intend to claim a resolution to the paradox of ``black hole information loss''. Rather, we have merely speculated on the qualitative features that a general quantum field theory for an arbitrary observer might possess. To consolidate our ideas about information loss, we would need a mathematical scheme for tracing out field degrees of freedom along the worldline of the observer and a corresponding detailed analysis. Moreover, we have adopted a semi-classical approximation, although we expect some features of the current theory, in particular the importance of causal structure on possible loss of information, to remain in a more complete theory. 

We have aimed at constructing a quantum theory for an arbitrary observer. To compare physical descriptions between arbitrary observers it is therefore very important to study quantum state transformation: specifically given a quantum state for an observer $\mathcal{C}^1(t)$ at time $t_0$, we would like to determine the corresponding state for another observer $\mathcal{C}^2(s)$ at time $s_0$. It is conceivable that when the simultaneity surface $\Sigma^2_{s_0}$ for observer $\mathcal{C}^2(s)$ is a subset of $\Sigma^1_{t_0}$ for observer $\mathcal{C}^1(t)$, the state transformation from $\mathcal{C}^1(t)$ at $t_0$ to $\mathcal{C}^2(s)$ at $s_0$ should be defineable. Furthermore, one would expect tracing out the extra degrees of freedom on $\Sigma^1_{t_0}$ to be necessary, so that a pure state for $\mathcal{C}^1(t)$ would in general be a mixed state for $\mathcal{C}^2(s)$. Finally, we require such a state transformation to be consistent with quantum state evolution. We will not however develop these any further. 

\section{Constructing the Dynamics of Quantum States}

We have constructed a one-parameter family of Fock spaces. The scalar function $t$ on which this construction depends is the time of each event for the observer defined in section 3, and the corresponding foliation of spacetime is the simultaneity surfaces of constant $t$. We have also argued that the vector space V of real solutions of Klein-Gordon equation should be defined on the diamond of the observer, which is the set of all points that can both receive light signals from and send light signals to the observer.

In this section, we turn to constructing the dynamics of quantum states for our theory: given a quantum state for an observer at his proper time $t_1$, what is the evolved quantum state at another time $t_2$? In other words, we need to construct a two-parameter mapping $U(t_2, t_1) : \ff(\hh_{t_1}) \to \ff(\hh_{t_2})$ such that:
\begin{enumerate}[(a)]
\item  $U(t_2, t_1)$ is an isomorphism, i.e. a complex-linear bijection; \label{a}
\item the inner product of $U(t_2,\, t_1)| \phi\rangle_{t_1}$ and $U(t_2,\, t_1)| \psi\rangle_{t_1}$ is equal to $\langle \phi|\psi \rangle_{t_1} $, for all\ $|\phi\rangle_{t_1}$ and $|\psi\rangle_{t_1}$ in $\ff(\hh_{t_1})$; and \label{b}
\item  $U(t_3, t_1) = U(t_3, t_2)\, U(t_2, t_1)$. \label{c}
\end{enumerate}
It then follows from these properties that $U(t,t) = \mathds{1}_t$ on $\ff(\hh_t)$ and that $U(t_1,t_2) = U^{-1}(t_2, t_1)$.

For notational simplicity, we will concentrate on some arbitrary choice of $t_1 = 0$ and $t_2 = t$. We use subscript $t$ on various symbols to denote its dependence on $t$ and/or $J_t$, and subscript $0$ for the case of $t=0$. In addition, we denote $U(t,\, 0)$ as $U_t$, although it should be understood that $U_t$ implicitly depends on a choice of $t_1=0$.

Consider a state of the form $C_0^+(\phi_1) \dots C_0^+(\phi_n)|0\rangle_0$, where $|0\rangle_0$ is the vacuum in $\ff(\hh_0)$. We define our $U_t$ by
\bb
U_t\,\Big( C_0^+(\phi_1) \dots C_0^+(\phi_n) \,|0\rangle_0 \Big) = \Big(U_t\, C_0^+(\phi_1)\Big) \dots \Big(U_t\, C_0^+(\phi_n)\Big)\, \Big( U_t\, |0\rangle_0\Big),
\label{3}
\ee
where we have used the same symbol $U_t$ to denote the mapping from $\ff(\hh_0)$ to $\ff(\hh_t)$ as well as the mapping from the complex vector space
\bb
\aa_0 := \{\alpha\, C_0^+(\phi) + \beta \, C_0^-(\psi):\ \alpha, \beta \, \in \mathbb{C},\ \phi, \psi \in V\}
\label{aaa}
\ee
to $\aa_t$ defined similarly to (\ref{aaa}). The nature of $U_t$, i.e. whether it is on $\ff(\hh_0)$ or $\aa_0$, should be clear from the argument on which it is acting.

Once we have defined $U_t$ on both $\aa_0$ and $|0\rangle_0$, generalising the action of $U_t$ to the entire Fock space $\ff(\hh_0)$ via complex-linearity and to density matrices should be straightforward. We shall then show that the mapping is well-defined and satisfies the three properties (\ref{a}), (\ref{b}), and (\ref{c}).\\\\

We now construct the map\footnote{We emphasise here that $C^\mu_0(\phi)$ and $C^\nu_t(\phi)$ are mappings on different Fock spaces and hence cannot be directly related by ``=''.} from $\aa_0$ to $\aa_t$. Our construction is guided by the following intuition: there is a ``natural'' change of creation/annihilation operators under a change of $J_t$; and there is a change of creation/annihilation operators induced by the evolution of their underlying classical fields.

We now capture the contribution of $J_t$ to $U_t$. Define
\bb
\aa_{\,t}^\mu := \{C^\mu_t(\phi):\ \phi \in V\},
\label{4.3}
\ee
where $\mu = \pm 1$. We have $\aa_{\,t} \cong \aa_{\,t}^+ \oplus \aa_{\,t}^-$, where  $\cong$ denotes isomorphism. The subspace $\aa_{\,t}^+$ is naturally isomorphic to $V_t$ via
\bb
C^+_t(\phi) \cong |\phi\rangle_t
\ee
where recall $V_t$ is the complex vector space constucted out of $(V,\, J_t)$ as in section 2. Similarly, the subspace $\aa^-_t$ is isomorphic to $\bar{V}_t$, the conjugate of $V_t$. 

Observe that there is also a natural isomorphism from $V_t$ to the $(+i)$ $J_t$-eigensubspace of $V^\mathbb{C} \cong V \oplus(iV)$. We shall denote this eigensubspace as $V^+_t$, and clearly $V^+_t=\{P^+_t \phi: \phi \in V\}$ where $P^+_t := (1 - iJ_t)/2\,$ is the projection from $V^\mathbb{C}$ onto $V^+_t$. This isomorphism is given by
\bb
|\phi\rangle_t \cong P^+_t\, \phi.
\ee
Similarly, $\bar{V}_t$ is isomorphic to $V^-_t$, the $(-i)$ eigensubspace of $V^\mathbb{C}$, via $P^-_J := (1+iJ_t)/2$. Above all, we have a natural isomorphism between $\aa_t$ and $V^\mathbb{C}$. We denote this isomorphism as $F_t$, that is:
\bb
F_t\,C^\mu_t(\phi) = P^\mu_t\, \phi,\qquad \forall \phi \in V,\quad \mu = \pm 1.
\label{4}
\ee

This is the equation we have been looking for, capturing the intuition that different $J_t$ corresponds to different notions of ``+'' and ``-'', embodied as different projections in our formalism. It is important to notice that $V^\mathbb{C}$ is independent of $J_t$. This disentangles the contribution of $J_t$ from the contribution of the classical evolution to $U_t$, which we now describe.

Let $u(t_2, t_1): V \to V$ be a two-parameter automorphism on $V$. We denote its complexification also by $u(t_2, t_1)$. It is obvious that this complexification defines an automorphism on $V^\mathbb{C}$. Again, we denote $u(t, 0)$ as $u_t$ for some arbitrary $t_1=0$, $t_2 = t$. 

To capture the contribution of $u_t$ to $U_t$, combined with that of $J_t$, we now define the map $U_t\!:\aa_0 \to \aa_t$ to be $U_t: = F_t^{-1} u_t\,F_0$:
\bb
\begin{array}{rcl}
\aa_0 & \xrightarrow{\quad  \textstyle{U_t} \quad } & \aa_t\\
\scriptstyle{F_0} \Big\downarrow & & \Big\uparrow \scriptstyle{F_t^{-1}} \\
V^\mathbb{C} & \xrightarrow{\quad \textstyle{u_t}\quad } & V^\mathbb{C}\\
\end{array}
\label{diagram}
\ee

Having defined $U_t$, we now determine the image of operators in $\aa_0$ under this map. Since $P^+_t\! +\! P^-_t$ is identity on $V^\mathbb{C}$:
\bb
\begin{array}{rl}
U_t\,C^\mu_0(\phi) &=F_t^{-1} (P^+_t\! +\! P^-_t)\, u_t\, F_0\; C^\mu_0(\phi)\\\\
&= F_t^{-1}(P^+_t u_t\, P^\mu_0\, \phi)+ F_t^{-1}(P^-_t u_t\, P^\mu_0\, \phi),\end{array}
\ee
where we used (\ref{4}). It is then straightforward to show that the action of $U_t$ on $\aa_0$ is
\bb
U_t\, C^\mu_0(\phi) = \sum_\alpha \, C^\alpha_t(\phi^\mu_{\alpha|t}),
\label{5}
\ee
where
\bb
\phi^\mu_{\alpha|t} = \tfrac{1}{2}(u_t - \mu\alpha\, J_t\, u_t J_0)\,\phi,
\label{soln}
\ee
with all Greek indices taking the value $\pm 1$. We shall henceforth drop the summation sign in any equation with repeated $\alpha$ or $\beta$ indices, with one index up and the other down. 

We have defined $U_t\!:\aa_0 \to \aa_t$ and determined its action on $\aa_0$ as given by (\ref{5}). To finish the construction of $U_t$ on $\ff(\hh_0)$, it remains to specify how vacuum evolves. Denote $|\chi\rangle_t$ to be the state evolved from vacuum: $|\chi\rangle_t = U_t\, |0\rangle_0$. We shall only impose that $|\chi\rangle_t$ is a uniquely defined unit vector in $\ff(\hh_t)$ and that it satisfies the condition
\bb
\Big(U_t\, C^-_0(\phi)\Big)|\chi\rangle_t = C^\alpha_t(\phi^-_{\alpha|t})\, |\chi\rangle_t = 0\qquad \forall\, \phi \in V.
\label{6}
\ee

Before specifying $|\chi\rangle_t$ and $u_t$, and before showing that our construction satisfies (\ref{a}), (\ref{b}), and (\ref{c}), we first say a few words about vacuum and particle creation. From (\ref{5}), we see that $C^+_0(\phi)$ will evolve to a pure creation operator if and only if $\phi^+_{-|t} = 0$ and that $C^-_0(\phi)$ will evolve to a pure annihilation operator if and only if $\phi^-_{+|t} = 0$. But $\phi^+_{-|t} = \phi^-_{+|t}$, so the two conditions are equivalent. If no $C^\mu_0(\phi)$ gets mixed, we have $(u_t + J_t\, u_t\, J_0)\phi = 0$ for all $\phi \in V$, i.e.
\bb
J_t = u_t\, J_0\, u_t^{-1}.
\label{7}
\ee
In this case, $U_t\, C^\mu_0(\phi) = C^\mu_t(u_t\,\phi)$. Additionally, (\ref{6}) reduces to
\bb
C^-_t(u_t\,\phi)|\chi\rangle_t = 0\qquad \forall \phi \in V.
\ee
The vector $|\chi\rangle_t$ is then fixed to be $|0\rangle_t$, the vacuum in $\ff(\hh_t)$, since $u_t$ is an automorphism on $V$. We then have
\bb
U_t|\phi_1\dots\phi_n\rangle_0 = |(u_t\phi_1)\,(u_t\phi_2)\dots (u_t\phi_n)\rangle_t.
\label{vac_evol2}
\ee
That is, a $n$-particle state in $\otimes^n_s\hh_0$ will evolve to a state in $\otimes^n_s\hh_t$ for all $n$: particles are not created. The mapping $U_t$ in (\ref{vac_evol2}) is indeed complex-linear, as illustrated by
\bb
\begin{array}{rl}
U_t\, i|\phi_1\phi_2\dots\phi_n\rangle_0 &= U_t |(J_0\phi_1)\,\phi_2\dots\phi_n\rangle_0\\
&= |(u_t\, J_0\, \phi_1)\,(u_t\,\phi_2)\dots (u_t\,\phi_n)\rangle_t\\
&= |(J_t\, u_t\, \phi_1)\,(u_t\,\phi_2)\dots (u_t\, \phi_n)\rangle_t\\
&= i\, |(u_t\, \phi_1)\,(u_t\,\phi_2)\dots (u_t\, \phi_n)\rangle_t\\
&= i\, U_t\, |\phi_1\,\phi_2\dots \phi_n\rangle_0
\end{array}
\ee
Therefore, we see that particle creation is precisely due to $J_t$ failing to evolve in a way satisfying (\ref{7}).\\\\

There are two conditions that the automorphism $u_t$ must satisfy. We now discuss the first condition motivated by condition (b) on $U_t$. To begin with, we have that
\bb
\langle \phi| \psi \rangle_0 = \langle 0|\,\, [C^-_0(\phi), C^+_0(\psi)] \,\,|0\rangle_0,\quad \textrm{for } |\phi\rangle_0,\, |\psi\rangle_0 \in \hh_0,
\label{inn_prod}
\ee
and that the inner product between $U_t| \phi \rangle_0$ and $U_t| \psi \rangle_0$ is given by
\bb
\langle \chi|\Big(U_t\, C^-_0(\phi)\Big) \Big(U_t\, C^+_0(\psi)\Big)| \chi \rangle_t,
\label{9}
\ee
where we have used the property
\bb
\Big(U_t\, C^\mu_0(\phi)\Big)^\dag = U_t\Big(C^\mu_0(\phi)\Big)^\dag.
\label{8}
\ee
To see this property, notice that the right-hand side of (\ref{8}) is
\bb
\begin{array}{rl}
  U_t\, C^{-\mu}_0(\phi) &= C^{-\alpha}_t(\phi^{-\mu}_{-\alpha|t})\\
    &= C^{-\alpha}_t(\phi^{\mu}_{\alpha|t})\\
    &= \Big(C^{\alpha}_t(\phi^{\mu}_{\alpha|t})\Big)^\dag,
\end{array}
\ee
where the final expression is simply the left-hand side of (\ref{8}). Condition (b) then implies
\bb
\langle 0|\,\, [C^-_0(\phi), C^+_0(\psi)] \,\,|0\rangle_0=\langle \chi|\Big(U_t\, C^-_0(\phi)\Big) \Big(U_t\, C^+_0(\psi)\Big)| \chi \rangle_t.
\label{star}
\ee
Using (\ref{6}), we can write the right-hand side of (\ref{star}) as
\bb
\langle \chi|\, \Big[U_t\, C^-_0(\phi), U_t\, C^+_0(\psi)\Big]\, | \chi \rangle_t.
\label{10}
\ee
Noticing that $\langle \chi| \chi \rangle_t = 1$ and that by (\ref{commutations}) the commutator in (\ref{10}) is a multiple of $\mathds{1}_t$ on $\ff(\hh_t)$, we see (\ref{star}) is equivalent to
\bb
[U_t\, C^-_0(\phi),\, U_t\, C^+_0(\psi)] = U_t\Big([C^-_0(\phi), C^+_0(\psi)]\Big)
\label{11}
\ee
where we have defined on the right-hand side
\bb
U_t\, \mathds{1}_0 = \mathds{1}_t.
\label{def1}
\ee
On the other hand, for (\ref{3}) to be well-defined, it is necessary that
\bb
[U_t\, C^+_0(\phi),\,\, U_t\, C^+_0(\psi)] = 0
\label{12}
\ee

That $U_t$ preserves the commutation relationship on all $\aa_0$ then follows from (\ref{8}), (\ref{11}), (\ref{12}), and complex-linearity. We now derive the condition on $U_t$ for this commutation to be preserved. A straightforward calculation using (\ref{inner_prod}), (\ref{symplectic1}), and (\ref{commutations}) shows that
\bb
[C^\mu_t(\phi), C^\nu_t(\psi)] = i\, \omega(P^\mu_t \, \phi,\, P^\nu_t \, \psi)\, \mathds{1}_t,
\label{13}
\ee
where we have extended the symplectic form $\omega$ given by (\ref{symplectic}) to $V^\mathbb{C}$ via complex-linearity. Applying (\ref{5}) and using the fact that both sides of (\ref{13}) are bilinear, we have
\bb
\begin{array}{rl}
[U_t\, C^\mu_0(\phi),\, U_t\, C^\nu_0(\psi)] &= \left[C^\alpha_t(\phi^\mu_{\alpha|t}), C^\beta_t(\psi^\nu_{\beta|t}) \right]\\
&= i\, \omega\left(P^\alpha_t \phi^\mu_{\alpha|t},\  P^\beta_t \psi^\nu_{\beta|t} \right)\mathds{1}_t.
\end{array}
\ee
Since 
\bb
\begin{array}{rl}
P^\alpha_t \, \phi^\mu_{\alpha|t} &= F_t\;C^\alpha_t(\phi^\mu_{\alpha|t})\, \\
&= F_t\,U_t\,C^\mu_0 (\phi) \\
&= u_t\,F_0\, C^\mu_0(\phi) \\
&= u_t\, P^\mu_0 \phi,
\end{array}
\ee
where we have used (\ref{4}), (\ref{diagram}), and (\ref{5}), we have
\bb
[U_t\, C^\mu_0(\phi),\, U_t\, C^\nu_0(\psi)] = i\, \omega\left(u_t\, P^\mu_0 \phi,\, u_t\, P^\nu_0 \psi \right)\mathds{1}_t.
\ee
Hence
\bb
[U_t\, C^\mu_0(\phi),\, U_t\, C^\nu_0(\psi)] = U_t\Big([C^\mu_0(\phi),\, C^\nu_0(\psi)]\Big)
\label{comm}
\ee
will be satisfied if and only if
\bb
\omega\,(u_t P^\mu_0\phi,\, u_t P^\nu_0\psi) = \omega\,(P^\mu_0\phi,\, P^\nu_0\psi).
\label{14}
\ee
Summing (\ref{14}) over $\mu,\, \nu$, one can show that this in turn is equivalent to
\bb
\omega (u_t\,\phi,\, u_t\, \psi) = \omega(\phi, \psi)\qquad \forall \, \phi, \psi \in V.
\label{15}
\ee
That is, $U_t$ preserves the commutation relation on $\aa_0$ if and only if $u_t$ preserves the symplectic form $\omega$ on $V$ (or equivalently on $V^\mathbb{C}$). With $u_t$ satisfying (\ref{15}), it is easy to generalise the proof that $U_t$ satisfies condition (b) for states in $\hh_0$ to the general case of all states in $\ff(\hh_0)$. Hence, in order for definition (\ref{3}) of $U_t$ to be well-defined and for $U_t$ to satisfy condition (\ref{b}), the first condition on $u_t$ which we impose is that it satisfies (\ref{15}).

The second condition, motivated by condition (\ref{c}), which we shall impose on $u_t$ is:
\bb
u(t_3,\, t_1) = u(t_3,\, t_2)\ u(t_2,\, t_1).
\label{17}
\ee\\\\

With a choice of automorphism $u_t$ on V satisfying (\ref{15}) and (\ref{17}), and a choice of $|\chi\rangle_t$ satisfying (\ref{6}), we shall now show our $U_t$ on $\ff(\hh_0)$ satisfies (\ref{a}), (\ref{b}), and (\ref{c}). 

First of all, we have shown (\ref{b}) is satisfied. This in particular implies $U_t$ is injective. Complex-linearity is obvious. Hence to show condition (\ref{a}) is satisfied, it remains to show that $U_t$ is onto.

We now show $U_t$ is onto. We first show there exists a $|\tilde{\chi}\rangle_0$ such that $U_t| \tilde{\chi} \rangle_0 = |0 \rangle_t$, the vacuum in $\ff(\hh_t)$. Let $|\tilde{\chi}\rangle_0$ be $|\tilde{\chi}\rangle_0 := U(0,\, t)|0\rangle_t$ where $U(0,\, t)$ is defined precisely in the same way as $U(t,\, 0)$ except for a change of ``$0$'' and ``$t$''. Then by construction using (\ref{6}), 
\bb
\Big(U(0,\, t)\, C^-_t(\phi)\Big)|\tilde{\chi} \rangle_0 = 0 \qquad \forall \phi \in V.
\label{4.33}
\ee
Since by using (\ref{6}) and (\ref{comm}) we have
\bb
U_t\Big( C^\alpha_0(\phi_\alpha)|\psi\rangle_0\Big) = \Big(U_t\, C_0^\alpha(\phi_\alpha)\Big)\,U_t |\psi\rangle_0 \qquad \forall \phi_\alpha \in V, \;\; |\psi\rangle_0 \in \ff(\hh_0),
\label{16}
\ee
we may apply $U_t$ on both sides of (\ref{4.33}) to obtain
\bb
\Big(U_t\, U(0,\, t)\, C^-_t(\phi)\Big)\Big(U_t |\tilde{\chi}\rangle_0\Big) = 0.
\label{18}
\ee
Observe that $U_t\, U(0,\, t)$ on $\aa_t$ is
\bb
F^{-1}_t u(t,\, 0) F_0\, F_0^{-1} u(0, t) F_t = \mathds{1}_t
\ee
where $u(t,\, 0)\, u(0,\, t) = \mathds{1}$ on $V$ follows\footnote{Note that we did not make any similar assumptions on $U_t$ for our proof.} from (\ref{17}) and the fact that $u(t_2,\, t_1)$ is an automorphism on $V$. Equation (\ref{18}) now becomes
\bb
C^-_t(\phi)\, (U_t |\tilde{\chi}\rangle_0) = 0 \qquad \forall \phi \in V
\ee
with a unique solution 
\bb
U_t|\tilde{\chi}\rangle_0 = |0\rangle_t.
\label{19}
\ee
We now conclude $U_t$ on $\ff(\hh_0)$ is onto, using (\ref{19}), (\ref{16}), and the fact that $U_t$ from $\aa_0$ to $\aa_t$ is an isomorphism. Furthermore on $\ff(\hh_t)$ we have
\bb
U(0,\, t) = U_t^{-1}.
\label{inv}
\ee

Finally, we prove that condition (\ref{c}) is satisfied. Applying $U(t^\prime,\, t)$ for an arbitrary $t^\prime$ to
\bb
\Big(U_t\, C^-_0(\phi)\Big)\, U_t |0\rangle_0 = 0 \qquad \forall \phi \in V
\ee
which is simply (\ref{6}), we have that
\bb
\Big(U(t^\prime,\, t)\, U_t\, C^-_0(\phi)\Big)\, U(t^\prime,\, t)\, U_t |0\rangle_0 = 0,
\ee
where we have used (\ref{16}) again. Observe that, using (\ref{diagram}),  the condition (\ref{17}) on $u_t$ is equivalent to condition (c) as a mapping on $\aa_0$. We have that
\bb
\Big(U_{t^\prime}\, C^-_0(\phi)\Big)\, U(t^\prime,\, t)\, U_t\, |0\rangle_0 = 0.
\label{20}
\ee
Apply $U(0,\, t^\prime) = U_{t^\prime}^{-1}$ to (\ref{20}) and use (\ref{16}) to obtain
\bb
C^-_0(\phi)\Big(U(0,\, t^\prime)\, U(t^\prime,\, t)\, U_t|0\rangle_0\Big)=0, \qquad \forall \phi \in V.
\ee
Hence,
\bb
U(0,\, t^\prime)\, U(t^\prime,\, t)\, U_t|0\rangle_0=|0\rangle_0.
\ee
Using (\ref{inv}) again, this is simply
\bb
U(t^\prime,\, t)\, U_t\,|0\rangle_0= U_{t^\prime} |0\rangle_0.
\label{21}
\ee
We finally conclude that condition (\ref{c}) is satisfied over all $\ff(\hh_0)$, using (\ref{21}), (\ref{16}), and the fact that condition (\ref{c}) is satisfied as mappings on $\aa_0$.\\\\

We now specify the evolution of vacuum from $|0\rangle_0$ to $|\chi\rangle_t $ satisfying (\ref{6}). To do so, we shall first extend the definition of the evolution map $U_t:\aa_0 \to \aa_t$ to sums of products of operators in $\aa_0$ via
\bb
U_t \Big(\prod_i C^{\mu_i}_0(\phi_i) \Big):=\prod_i U_t \, C^{\mu_i}_0(\phi_i) 
\label{def}
\ee
and complex linearity. Clearly this mapping is well defined, following from definition (\ref{def1}) and that $U_t$ is linear on $\aa_0$ and preserves commutation (\ref{comm}). Furthermore, this extended $U_t$ is invertible, satisfies the desired composition as in condition (c) at the beginning of this chapter, and commutes with taking the adjoint of its argument as in (\ref{8}). 

The evolution of vacuum is then given by applying (\ref{def}). Observe that the vacuum density matrix can \emph{formally} be written as
\bb
|0\rangle \! \langle 0|_{\,0} = \prod_{n=1}^\infty (\mathds{1}_0 - N_0/n)
\label{formal}
\ee
where the total number operator $N_0$ on $\ff(\hh_0)$ is defined by $N_0|\phi^n\rangle_0 := n|\phi^n\rangle_0$ for all $|\phi^n\rangle_0 \in \otimes^n_s\hh_0$. This can be seen by noticing that $\langle 0 |0\rangle \! \langle 0| 0 \rangle_{0} = 1$ and $\langle \psi^m |0\rangle \! \langle 0| \phi^n \rangle_{0} = 0$ are true for both sides of (\ref{formal}), for $|\phi^n\rangle_0 \in \otimes^n_s\hh_0$ and $|\psi^m\rangle_0 \in \otimes^m_s\hh_0$ with at least one of $m$ or $n$ being non-zero. The evolved state is then given by
\bb
\rho_t:=U_t\big(|0\rangle \! \langle 0|_{\,0}\big) = \prod_{n=1}^\infty \Big(\mathds{1}_t - (U_t\,N_0)/n\Big).
\label{vac_evol}
\ee
The operator $U_t\,N_0$ is also defined using (\ref{def}). We write $N_0$ as
\bb
N_0 = \sum_k C^+_0(\phi_k) C^-_0(\phi_k)
\ee
where summation is over all $k$ such that $\{|\phi_k\rangle_0\}$ is an orthonormal basis for $\hh_0$. Then we have that 
\bb
U_t\,N_0 = \sum_k \Big(U_t\,C^+_0(\phi_k)\Big)\Big(U_t\, C^-_0(\phi_k)\Big).
\ee
One can show that this definition of $U_t\,N_0$ is independent of a choice of an orthonormal basis $\{|\phi_k\rangle_0\}$ for $\hh_0$.

In general, (\ref{vac_evol}) does not define a finite-trace operator on $\ff(\hh_t)$, in which case strictly speaking our construction of unitary dynamics fails mathematically, although physically meaningful predictions can still be made. This, our only problem with infinity, resembles the renormalisation problem in the usual field theory of quantum operators. In a more complete theory, the dimension of the physical Hilbert space should probably be replaced by a finite one, so we believe this difficulty does not indicate that our formulation is fundamentally towards a wrong direction.

In the case when (\ref{vac_evol}) does converge to a finite-trace operator, however, we shall show that $\rho_t$ defined in (\ref{vac_evol}) is a pure density matrix. First of all, (\ref{def}) maps hermitian operators to hermitian operators, using (\ref{8}). It then follows $\rho_t^{\dag}=\rho_t$. Secondly, we have $\rho_t^2=\rho_t$, since
\bb
U_t \big( |0\rangle \! \langle 0|_{\,0} \big) \,U_t \big( |0\rangle \! \langle 0|_{\,0} \big)=U_t \big( |0\rangle \! \langle 0|\,|0\rangle \! \langle 0|_{\,0} \big)=U_t \big( |0\rangle \! \langle 0|_{\,0} \big)
\ee
where we have used (\ref{def}). It follows from these two properties that $\rho_t$ is a projector with eigenvalues 1 or 0, so that we can write $\rho_t$ in an orthonormal basis as
\bb
\rho_t:= \sum_{k=1}^n |k\rangle \! \langle k|
\ee
where $n$ is the trace of $\rho_t$ if it is finite. We shall now show in this case $n=1$ and that $\rho_t$ defined in (\ref{vac_evol}) is a pure density matrix. Evolving the identity $\rho_t |i\rangle \! \langle i|=|i\rangle \! \langle i|$ for an arbitrary $i$ between 1 and $n$ by $U(0,t)$ we obtain
\bb
|0\rangle \! \langle 0|_{\,0}\;\; U(0,t)\big( |i\rangle \! \langle i| \big)=U(0,t)\big( |i\rangle \! \langle i| \big)
\ee
where we have used (\ref{def}). Similarly,
\bb
U(0,t)\big( |i\rangle \! \langle i| \big) \;\;  |0\rangle \! \langle 0|_{\,0}=U(0,t)\big( |i\rangle \! \langle i| \big)
\ee
Hence $U(0,t)\big( |i\rangle \! \langle i| \big)$ is proportional to $|0\rangle \! \langle 0|_{\,0}$, that is, $ |i\rangle \! \langle i|$ is proportional to $U_t |0\rangle \! \langle 0|_{\,0}$ for all $i$. This is only possible if $n=1$, i.e. $\rho_t=|\chi\rangle\! \langle \chi|_t$ for some pure state $|\chi\rangle_t$ as desired.

We have specified the evolution of vacuum as in (\ref{vac_evol}), and we have demonstrated that it indeed defines a pure state $|\chi\rangle_t$.  It remains to check that $|\chi\rangle_t$ satisfies condition (\ref{6}), i.e. $ C^\alpha_t(\phi^-_{\alpha|t})\, |\chi\rangle_t = 0$. This is the case if and only if 
\bb
C^\alpha_t(\phi^-_{\alpha|t})\, |\chi\rangle\! \langle \chi|_t \,  C^{-\beta}_t(\phi^-_{\beta|t}) = 0.
\ee
But by (\ref{def}) the left hand side of this is
\bb
U_t C^-_0(\phi) \,U_t\big(|0\rangle \! \langle 0|_{\,0}\big)\, U_t C^+_0(\phi)=U_t \Big( C^-_0(\phi) |0\rangle \! \langle 0|_{\,0} C^+_0(\phi) \Big)
\ee
which is simply 0 as desired. 

We say a final word about particle creation. If the quantum state was vacuum at time zero, then the expected particle number at time $t$ is $\textrm{tr}(|\chi\rangle\!\langle \chi|_t\,N_t)$, where $N_t$ is the total number operator for $\ff(\hh_t)$ defined in the same way as $N_0$ for $\ff(\hh_0)$. For the $|\phi\rangle_t $ particle number operator $C^+_t(\phi) C^-_t(\phi)$, using that $U_t$ is trace-preserving, we have 
\bb
\textrm{tr}\Big(|\chi\rangle\!\langle \chi|_t\, C^+_t(\phi) C^-_t(\phi) \Big) = \textrm{tr} \Big( |0\rangle \! \langle 0 |_0\, \Big(U(0,t)\, C^+_t(\phi) C^-_t(\phi) \Big) \Big).
\ee
It is then straightforward to calculate that the expected $|\phi\rangle_t $ particle number seen by the observer at his proper time $t$ is given by the square of the norm of $|\phi^-_{+|0}\rangle_0$, where
\bb
\phi^-_{+|0} := \tfrac{1}{2}\, \big(\, u_t^{-1}+J_0 \,u_t^{-1} J_t \big)\,\phi.
\ee
See (\ref{soln}) for a comparison of definitions.\\\\

Finally it remains to specify the classical evolution $u_t$, the automorphism on $V$, satisfying (\ref{15}) and (\ref{17}). One natural choice seems to be the following. For each $\phi \in V$, we define $u_t\,\phi \in V$ to be the unique solution with Cauchy data\footnote{Here we have included a factor of $\sqrt{h}$ in the definition of Cauchy data which is slightly different from the convention in section 2.} on $\Sigma_t$
\bb
\begin{array}{rl}
u_t\, \phi|_{\Sigma_t} =& \phi|_{\Sigma_0}\\
\sqrt{h_t}\,n^a\nabla_a(u_t\phi)|_{\Sigma_t} =& \sqrt{h_0}\,n^a\nabla_a\phi|_{\Sigma_0} 
\end{array}
\ee
where we have used a natural point identification map from $\Sigma_0$ to $\Sigma_t$ given by the integral curves of $\partial/\partial t$. Clearly this mapping is defined independent of the labelling of the 3-parameter family of  the integral curves of $\partial/\partial t$. It can be shown that it indeed is an automorphism on $V$ satisfying (\ref{15}) and (\ref{17}). However, it remains future work to check whether this choice of $u_t$, although self-consistent, leads to the same physical conclusions as those in the various well-known cases. 

[Aside. One may attempt to define the classical evolution by ``pulling'' Cauchy data back from $\Sigma_t$ to $\Sigma_0$ along $\partial / \partial t$, instead of ``pushing'' Cauchy data as above. More precisely, one may define $\tilde{u_t}$ such that
\bb
\begin{array}{rl}
\tilde{u}_t\, \phi|_{\Sigma_0} =& \phi|_{\Sigma_t}\\
\sqrt{h_0}\,n^a\nabla_a(\tilde{u}_t\phi)|_{\Sigma_0} =& \sqrt{h_t}\,n^a\nabla_a\phi|_{\Sigma_t}.\\
\end{array}
\ee
It can be shown that $\tilde{u_t}$ is also an automorphism on $V$ and satisfies (\ref{15}). However, it does not satisfy (\ref{17}). In fact $\tilde{u}_t = u^{-1}_t$, hence $u(t^\prime,\,t)\,u_t = u_{t^\prime}$ implies that $\tilde{u}_t\,\tilde{u}(t^\prime,\,t) = \tilde{u}_{t^\prime}$, which is not the physically desired composition law.]

\section{Summary and Outlook}

We summarise the main results of our discussion here. 

The goal of this paper is to propose a quantum field theory for an arbitrary observer in curved spacetime. To this end, we constructed a one-parameter family of Fock spaces based on the formalism of Ashtekar and Magnon \cc{am}. Each Fock space is based on a Hilbert space constructed from the vector space $V$ of real-valued Klein-Gordon solutions and a parameter-dependent complex structure $J_t$. This construction requires a choice of scalar function $t$ such that each constant $t$ hypersurface $\Sigma_t$ is a spacelike Cauchy surface.

We then applied this mathematical formalism for an arbitrary observer to the region of spacetime which the observer can both send signals to and receive signals from. Following \cc{diamond}, we have used the terminology ``diamond'' to refer to this region. We argue that radar time should be applied for the above function $t$ used in the construction of the Fock spaces. Physically this means $\Sigma_t$, the set of points in the diamond with radar time $t$, is the set of all events ``simultaneous'' to the observer at his proper time $t$. Our definition using light signal communication has a clear operational meaning and directly relates each point to the observer's local physical quantities. Our slicing applies to all observers and reflects the causal structure of the underlying spacetime.  

Although the diamond of an observer in general may not cover the entire spacetime, it is operationally the only region physically relevant to the observer. We therefore define the vector space $V$ of real Klein-Gordon solutions and our quantum field theory on the diamond of a given observer. We regard this as an axiom of our framework. 
 
In the case where all simultaneity surfaces of an observer are Cauchy surfaces of his diamond, we have constructed a unitary dynamics where no information is lost: given a quantum state for an observer at his proper time $t_1$, we constructed a two-parameter mapping $U(t_2, t_1)$ from $\ff(\hh_{t_1})$ to $\ff(\hh_{t_2})$ that will give us the evolved state at time $t_2$. We require our mapping to satisfy three conditions: that $U_t$ is an isomorphism; that the mapping preserves the inner product; and that the mapping satisfies $U(t_3, t_2)\,U(t_2, t_1) = U(t_3, t_1)$. 

The action of $U_t$ on $\ff(\hh_0)$ is specified by its action on the vector space $\aa_0$ of creation and annihilation operators given by (\ref{aaa}), as well as its action on the vacuum in $\ff(\hh_0)$. The construction is guided by the intuition that there is an evolution of operators in $\aa_0$ according to a change in the complex structure $J_0$, as well as according to an evolution $u_t$ of the operators' underlying classical fields, i.e. an automorphism on $V^\mathbb{C}$. The evolution of vacuum state is given by (\ref{vac_evol}) and satisfies (\ref{6}). We have deduced that particle creation will take place if and only if $J_t$ fails to evolve according to $J_t = u_t\, J_0\, u_t^{-1}$. 

We have also speculated on features that the theory covering the more general cases might have. We have further developed Wald's insight \cc{wald} that there should generally be a loss of information when there is evolution of a Cauchy surface to a non-Cauchy surface. Our one-parameter foliation of the diamond gives this ``evolution of surfaces'' a precise mathematical meaning. Physically, we give these ``surfaces'' an operational meaning as the simultaneity surfaces of a given observer. Moreover, the question of whether or not a surface is a Cauchy surface is now addressed with respect to the observer's diamond. 

Above all, we speculate that information will be lost precisely whenever an observer's surface of simultaneity evolves from a Cauchy surface to a non-Cauchy surface, whether this be due to the background spacetime or to the observer's motion in that background. This generalises Hawking's original insight \cc{Hawking:1976, Hawking:1982} and takes the information loss of black hole evaporation as just a special case.\\\\

There is much scope for future generalisations and applications of this work.

\begin{enumerate}[\textbullet]
\item We would like to generalise our current formulation to the cases when not all surfaces of simultaneity of an observer are Cauchy surfaces for his diamond. As we evolve data from a Cauchy surface to a non-Cauchy surface, we need a general scheme to trace out the field degrees of freedom lost along the worldline of the observer. Such a scheme would also consolidate our speculations about information loss.

\item It is important to apply our framework to various examples, especially to see how our results would compare with experiments and with the results obtained by other formulations, such as with canonical quantization or with model particle detectors.\\
In particular, it would be very important to explore quantum field theory in FRW cosmology using our formalism, where the simultaneity surfaces for a co-moving observer are not the usual surfaces of constant energy density. This offers the opportunity to test our theory with cosmological observations. 

\item We would like to generalise our formalism to higher-spin fields and interacting fields.

\item One important extension of our work is to study quantum state transformation between arbitrary observers and its operational meaning. Given a quantum state for an observer $\mathcal{C}^1(t)$ at time $t_0$, we would like to know the corresponding state for another observer $\mathcal{C}^2(s)$ at time $s_0$. We require such a transformation to be consistent with the state evolution we have already constructed.

\item We constructed a unitary dynamics relating a quantum state at some time of an observer to a state at a later time. However we have not discussed, for an arbitrary observer in curved spacetime, how a quantum measurment projects quantum states nor how the notion of ``wavefunction collapse'' should be understood.

\item One of the most intriguing aspects of quantum field theory in curved spacetime is the yet to be understood deep relationship between causal horizons and thermodynamics. Firstly, it has been argued \cc{JP} that the ultimate significance of the thermodynamics of black hole horizons hangs on the issue of its generalisation. In our framework, the diamond for an arbitrary observer provides a natural and most general notion of causal horizons, and by defining quantum theory over this region, our formalism naturally establishes a link between quantum theory and causal horizons. 

Secondly, thermodynamics of causal horizons can be studied using a statistical mechanics approach. This requires a concept of particles, and our formulation indeed provides a notion of particles for each observer. On the other hand, it would also be very enlightening from our formalism to obtain an expression directly relating thermodynamic quantities and properties of causal horizons without needing to consider any particle spectrum.

Finally, it has been argued that entropy is an observer-dependent quantity \cc{Marolf}. Our observer-dependent quantum field theory may be an ideal framework for a further investigation of the observer-dependence of thermodynamic quantities.   

\item We propose that observer-dependence as a fundamental feature of quantum field theory should be taken much further. Since the quantum states of particles in general depend on the observer, therefore different observers will have different $\langle T_{\mu\nu} \rangle$, the expectation values of the energy-momentum associated with their respective quantum states. By Einstein's field equations, the quantum field's back-reaction on the spacetime metric will also be observer-dependent and hence so will the spacetime metric itself. Indeed, the possibility that spacetime itself may be observer-dependent has been suggested by Gibbons and Hawking \cc{GH}. Such a profound suggestion merits further investigation which our framework may be well-suited to pursue, and this investigation may indeed be the correct path leading to a complete and consistent union of quantum theory and relativity.

\end{enumerate}

\section*{Acknowledgements}
\addcontentsline{toc}{section}{Acknowledgments}

I would like to thank Rex Liu for discussions, David Wiltshire and Steffen Gielen for comments and Jonathan Oppenheim for criticisms.
                                          


\end{document}